\pdfoutput=1
\documentclass[prx,twocolumn,amsmath,amssymb,floatfix]{revtex4-1}

\usepackage{graphicx}     
\usepackage{bm}        
\usepackage[colorlinks = true,
linkcolor = blue,
urlcolor  = blue,
citecolor = blue,
anchorcolor = blue]{hyperref}
\usepackage{url} 
\usepackage{dcolumn}   
\usepackage{mathrsfs}
\usepackage{epigraph}
\usepackage{braket}
\usepackage{gensymb}
\usepackage{lmodern}
\usepackage{lipsum}
\usepackage{marvosym}
\usepackage{ tipa }
\usepackage{bbold}
\usepackage{dsfont}
\usepackage{esint}
\usepackage{mathdots}
\usepackage{xcolor}
\usepackage{appendix}
\usepackage{natbib}
\usepackage{multirow}
\begin{document}
	
\title{When can localized spins interacting with conduction electrons in ferro- or antiferromagnets be described classically via the Landau-Lifshitz equation: Transition from quantum many-body entangled to quantum-classical nonequilibrium states}	
\author{Priyanka Mondal}
\affiliation{Department of Physics and Astronomy, University of Delaware, Newark, DE 19716, USA}
\author{Abhin Suresh}
\affiliation{Department of Physics and Astronomy, University of Delaware, Newark, DE 19716, USA}
\author{Branislav K. Nikoli\'{c}}
\email{bnikolic@udel.edu}
\affiliation{Department of Physics and Astronomy, University of Delaware, Newark, DE 19716, USA}
		
\begin{abstract}
Experiments in spintronics and magnonics operate with macroscopically large number of localized spins within ferromagnetic (F)  or antiferromagnetic (AF)  materials, so that their nonequilibrium dynamics is standardly described by the Landau-Lifshitz (LL) equation treating localized spins as {\em classical vectors of fixed length}. However, spin is a genuine quantum degree of freedom, and even though quantum effects become progressively less important for spin value $S>1$, they exist for all $S < \infty$. While this has motivated exploration of limitations/breakdown of the LL equation---by using examples of F insulators and comparing LL trajectories to quantum expectation values of localized spin operators---analogous comparison of fully quantum many-body vs. quantum (for electrons)-classical (for localized spins) dynamics in systems where {\em nonequilibrium} conduction electrons are present is lacking. Here we employ quantum Heisenberg F or AF chains of $N=4$ sites, whose localized spins interact with conduction electrons via $sd$ exchange interaction,  to perform such comparison by starting from {\em unentangled} pure (at zero temperature) or mixed (at finite temperature) quantum state of localized spins as the initial condition. This reveals that quantum-classical dynamics can faithfully reproduce fully quantum dynamics in the F metallic case, but only when spin $S$, Heisenberg exchange between localized spins and $sd$ exchange are sufficiently small. Increasing any of these three parameters can lead to substantial deviations, which are explained by the {\em dynamical buildup} of  entanglement between localized spins and/or between them and electrons. In the AF metallic case, substantial deviations appear even at early times, despite starting from {\em unentangled} N\'{e}el state, which therefore  poses a challenge on how to rigorously justify wide usage of the LL equation in phenomenological modeling of antiferromagnetic spintronics experiments. We also discuss finite temperature and finite size effects to demonstrate that: ({\em i}) including thermal fluctuations delays the onset of dynamical buildup of entanglement, but it does not suppress it; and ({\em ii}) no significant changes in the dynamics of this particular problem occur as we increase the chain length to $N>4$ sites, while even $N=4$ is sufficient to observe quantum-chaotic energy level-statistics whose emergence in such a small quantum many-body system crucially relies on interaction between localized spins and conduction electrons.

\end{abstract}
\maketitle

\section{Introduction}\label{sec:intro}

The development of consistent schemes for describing interaction of quantum and classical systems has been an 
interdisciplinary effort~\cite{Berry1993,Zhang2006,Agostini2007,Elze2012,Radonjic2012,Barcelo2012,Oliynyk2016} due to great interest from both conceptual viewpoint and demands of practical calculations in physics and chemistry. Regarding the former, Copenhagen interpretation~\cite{Rovelli1996} of quantum mechanics treats measuring devices  as classical objects which interact with a quantum system of interest~\cite{Nikolic2015}. Regarding the latter, in, e.g., chemical physics and chemistry it is impossible to numerically solve the Schr\"{o}dinger equation for a macroscopically large number of particles. This has lead to a plethora of hybrid quantum-classical schemes~\cite{Kapral2005,Kapral2015} where  heavy particles  (such as ions, atoms and molecules) are treated by classical molecular dynamics (MD)  and only a much smaller number of  lighter particles (such as electrons and protons~\cite{Markland2018,Kantorovich2018}) requires quantum dynamics that is (possibly nonadiabatically~\cite{Kantorovich2018,Agostini2019,Dou2018}) coupled to classical MD. Other examples that have motivated development of quantum-classical hybrid approaches include electrons interacting with nanoelectromechanical systems~\cite{Bode2011,Thomas2012,Lu2012} or fast electronic spins interacting with slow classical local magnetization $\mathbf{M}_i$ of magnetic materials~\cite{Ralph2008,Baltz2018} embedded into spintronic devices~\cite{Bode2012,Petrovic2018,Bajpai2019a,Bajpai2020,Petrovic2021b,Sayad2015,Bostrom2019,Elbracht2020}. In both of these cases, electrons comprise a quantum system which is  {\em open and out of equilibrium}.

\begin{figure}
	\centering
	\includegraphics[scale=0.3]{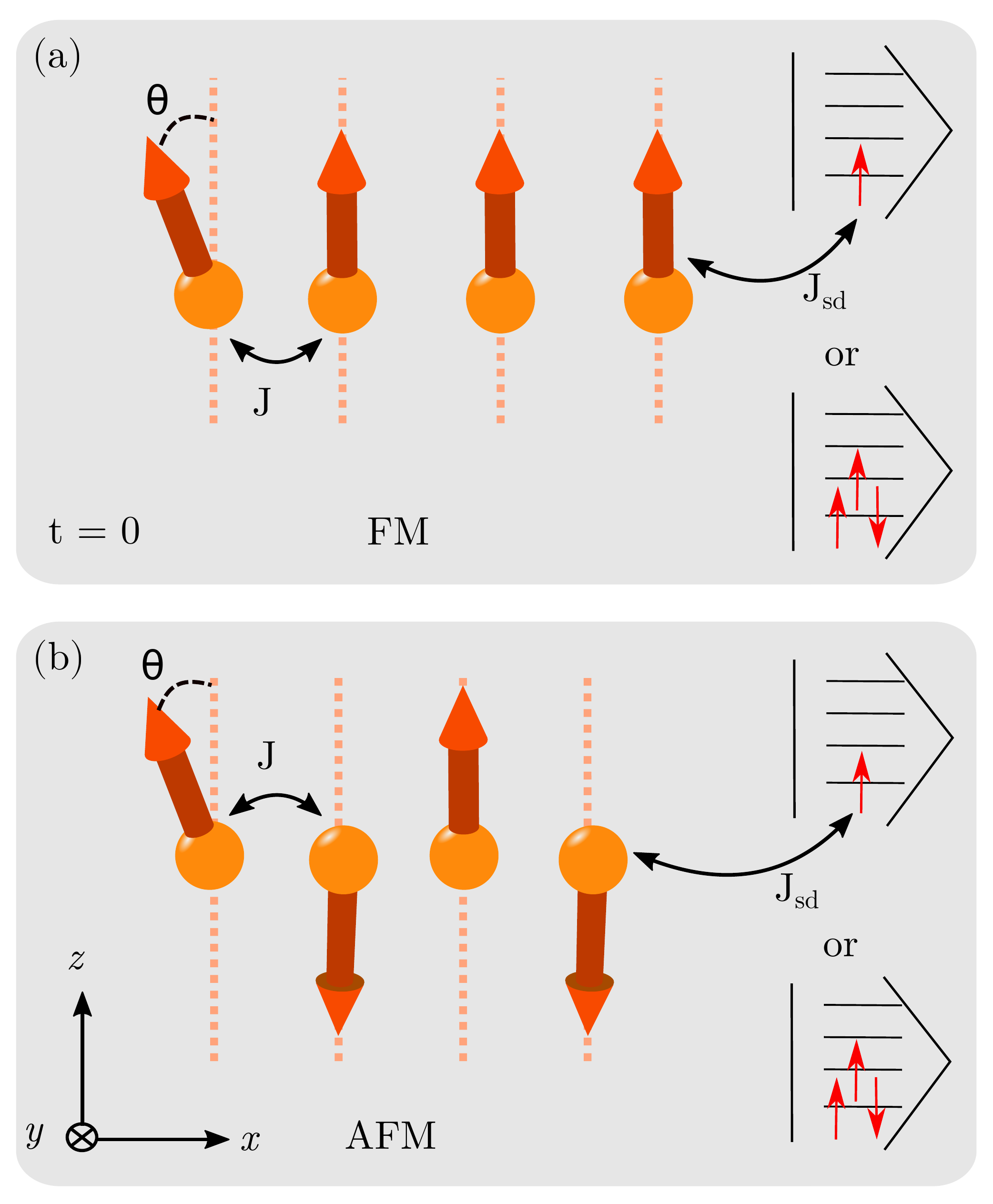}
	\caption{Schematic view of 1D chains of $N=4$ sites $i$  hosting quantum $\hat{\mathbf{S}}_i$ or classical $\mathbf{S}_i$ localized spins which interact  via (a) NN ferromagnetic $J>0$ or (b) NN antiferromagnetic $J<0$ Heisenberg exchange. Conduction electrons hop between the sites, making 1D chains metallic, while their movable spins interact with localized spins via $sd$ exchange interaction of strength $J_\mathrm{sd}$. We use $N_e=1$ or $N_e=3$ electrons to fill the chain with their kets in the Fock space illustrated in the insets. The nonequilibrium dynamics is initiated~\cite{Wieser2015} by applying magnetic field locally at site $i=1$, $B_1^x(t \ge 0) \neq 0$. To enforce {\em unentangled} N\'{e}el state  $|\uparrow \downarrow \ldots \uparrow \downarrow \rangle$ as the GS for $t \le 0$ of AFM, thereby making possible~\cite{Wieser2015} comparison with classical [via the LL Eq.~\eqref{eq:llg}] modeling of localized spins, we employ the usual trick~\cite{Stoudenmire2012,Mitrofanov2021} of staggered magnetic field $B_i^z=-B_{i+1}^z$.}
	\label{fig:fig1}
\end{figure}

Although a number studies have been focused on the general structure and properties of quantum-classical hybrid schemes, none are regarded as being the definitive solution. In the case of quantum electrons interacting with classical localized spins, rigorous derivation of quantum-classical hybrid is bypassed   
and one instead relies on the celebrated Landau-Lifshitz (LL) equation 
\begin{eqnarray}\label{eq:llg}
	\frac{\partial \mathbf{S}_i}{\partial t} & = & -g \mathbf{S}_i \times \mathbf{B}^\mathrm{eff}_i + \lambda \mathbf{S}_i \times \frac{\partial \mathbf{S}_i}{\partial t},   
\end{eqnarray}
introduced in 1935~\cite{Landau1935} as a purely phenomenological one to treat localized spins $\mathbf{S}_i$ (or, equivalently, $\mathbf{M}_i \propto \mathbf{S}_i$) as classical vectors of fixed length ($|{\bf S}| = 1$) which can, therefore, {\em only rotate}. Here $g$ is the gyromagnetic ratio; $\mathbf{B}^{\rm eff}_i = - \frac{1}{\mu_M} \partial H_\mathrm{lspins} /\partial \mathbf{S}_i$ is the effective magnetic field, as the sum of external field, field due to interaction with other localized spins and magnetic anisotropy in the classical Hamiltonian of localized spins $H_\mathrm{lspins}$; and $\mu_M$ is the magnitude of magnetic moments associated with localized spins~\cite{Evans2014}.  The LL equation has become the cornerstone of classical micromagnetics~\cite{Berkov2008,Kim2010} and atomistic spin dynamics~\cite{Evans2014} numerical schemes that are widely used to interpret~\cite{Woo2017} experiments in contemporary spintronics and magnonics. The precessional term [first on the right-hand side in Eq.~\eqref{eq:llg}] can be derived from quantum mechanics~\cite{Wieser2011}. The damping term (second on the right-hand side in Eq.~\eqref{eq:llg} and written originally~\cite{Landau1935} in a different ``LL form''~\cite{Saslow2009}) was justified by Gilbert~\cite{Saslow2009} using the Lagrange formalism with the classical Rayleigh damping---it can also be derived from quantum mechanics, but by starting~\cite{Wieser2015} from a non-Hermitian Hamiltonian operator in order to take into account energy dissipation~\cite{Saslow2009}. The Gilbert damping parameter $\lambda$ is inserted ``by hand'' with value taken from experiments~\cite{Weindler2014}  or from quantum calculations~\cite{Starikov2010} employing single-electron Hamiltonian with spin-orbit or magnetic disorder scattering. Additional steady-state~\cite{Ellis2017} of time-dependent~\cite{Petrovic2018,Bostrom2019} quantum transport calculations are also often used to 
add microscopic (rather than  phenomenological~\cite{Berkov2008,Kim2010,Evans2014}) Slonczewski-Berger spin-transfer torque~\cite{Ralph2008} (STT) term 
due to injected electronic spin current or as back-action~\cite{Bajpai2019a,Bajpai2020,Petrovic2021b,Sayad2015,Elbracht2020} of electrons driven out of equilibrium by $\mathbf{S}_i(t)$ itself.   

In all quantum derivations of the LL equation, one constructs the Heisenberg equation of motion for the spin operators  $\hat{\mathbf{S}}_i$, whose expectation values are then replaced (in the spirit of the Ehrenfest theorem,  which states that the expectation values obey the classical dynamical laws~\cite{Ballentine1994}) by classical vectors to finally arrive~\cite{Wieser2011,Wieser2015} at Eq.~\eqref{eq:llg}. Although spin is a genuine internal quantum degree of freedom, a na\"{i}ve~\cite{Goldberg2020} ``rule of thumb'' is that spin dynamics can transition to becoming  classical and, therefore, amenable to modeling by the LL equation in the limit $S \rightarrow \infty$ and $\hbar \rightarrow 0$ (while $S\hbar \rightarrow 1$). This is motivated by, e.g., the  eigenvalue of $\hat{\mathbf{S}}_i^2$ being $S^2(1+1/S)$, instead of $S^2$, which suggests that quantum effects become progressively less important for $S>1$.  However, they actually persist for all $S < \infty$, vanishing as $(2S)^{-1}$ in the classical limit~\cite{Parkinson1985}. Furthermore, in contrast to chemical physics and chemistry where many nuclei are much heavier than electrons so that classical MD safely applies to them~\cite{Markland2018}, most of standard magnetic materials host localized spins with rather small $S \le 5/2$~\cite{Kaxiras2019}. 

This issue has ignited intense  efforts~\cite{Wieser2011,Wieser2015,Wieser2013,Gauyacq2014} to delineate limitations/breakdown of the LL equation by directly comparing quantum $\langle \hat{\mathbf{S}}_i \rangle(t)$ vs. classical $\mathbf{S}_i(t)$ [computed by LL equation with $\lambda=0$ in Eq.~\eqref{eq:llg}]  trajectories and their Fourier transforms~\cite{Gauyacq2014} for systems of localized spins {\em alone} in models of ferromagnetic (F) insulators. Such studies for a {\em single} localized spin have revealed~\cite{Wieser2011,Gauyacq2014} that the presence of terms quadratic in $\hat{\mathbf{S}}_i$, such as magnetic anisotropy $D_z(\hat{S}_i^z)^2$ or biquadratic exchange $J_b(\hat{\mathbf{S}}_i \cdot \hat{\mathbf{S}}_j)^2$~\cite{Kartsev2020}, makes quantum and classical trajectories substantially different for small $S$. Although the discrepancy can be reduced by increasing $S$ to very large values~\cite{Gauyacq2014}, this works only for finite time interval (which elongates with $S \rightarrow \infty$) because of always present quantal revivals~\cite{Gauyacq2014}. This finding is not surprising in the light of the breakdown~\cite{Ballentine1994} of the Ehrenfest theorem for any nonlinear potential $V(\hat{\mathbf{r}})$ because $\langle \hat{\mathbf{r}}^p \rangle \neq \langle \hat{\mathbf{r}} \rangle^p$ when $p \ge 2$.  
		
For {\em many} localized spins, a more demanding requirement for applicability of coupled LL equations is that quantum many-body state of localized spins {\em must}~\cite{Wieser2015} remain {\em unentangled or separable}~\cite{Chiara2018,Bardarson2012,Petrovic2021a}, \mbox{$|\Sigma (t) \rangle_\mathrm{lspins} = |\sigma_1(t) \rangle \otimes |\sigma_2(t) \rangle \otimes \cdots |\sigma_N(t) \rangle$}, for all times $t$. Otherwise, if one  starts from unentangled states, such as the ground state (GS) of a ferromagnet \mbox{$|\mathrm{GS}\rangle_\mathrm{F}=| \uparrow \uparrow \ldots \uparrow \uparrow \rangle$}, but their superpositions and thereby entanglement are created dynamically~\cite{Petrovic2021a}, one finds that vector magnitude $|\langle  \hat{\mathbf{S}}_i \rangle (t)|  < S\hbar$ is shrinking (sometimes even to zero~\cite{Petrovic2021a}) as the {\em entanglement} grows in time. Entanglement~\cite{Chiara2018} describes genuinely quantum and nonlocal correlations between different parts of a physical system. Thus, it {\em cannot} be mimicked by any classical equation~\cite{Wieser2015}. 

Despite these insights, no study has compared fully quantum many-body vs. quantum-classical nonequilibrium dynamics for a system of  localized spins interacting with conduction electrons. This is largely due to the fact that self-consistent quantum-classical schemes have been developed only very recently~\cite{Petrovic2018,Bajpai2019a,Bajpai2020,Petrovic2021b,Sayad2015,Bostrom2019,Elbracht2020}, in contrast to earlier efforts~\cite{Onoda2006,Nunez2008,Fransson2008,Hurst2020} attempting to ``integrate  out'' fast electrons and arrive at an extended LL equation for localized spins only---a strategy which can never succeed~\cite{Bajpai2019a,Sayad2015} in all relevant parameter regimes.

Furthermore, in the case of both antiferromagnetic (AF) insulators and metals, the true GS is always substantially different from unentangled N\'{e}el state $|\Sigma\rangle^{\rm N\acute{e}el}_\mathrm{lspins}=| \uparrow \downarrow \ldots \uparrow \downarrow \rangle$ assumed as the initial state 
when employing classical atomistic spin dynamics codes~\cite{Evans2014}. Its nonzero entanglement is present even in equilibrium~\cite{Christensen2007,Kamra2019,Petrovic2021} (below some ``entanglement temperature''~\cite{Mathew2020,Scheie2021}). For example, in a minimal system [Fig.~\ref{fig:fig1}(b) but without any conduction electrons]  of four localized spins $S=1/2$, described by the quantum Heisenberg Hamiltonian $\hat{H}_\mathrm{AF}$ with the nearest-neighbor (NN) exchange interaction $J<0$, the true GS is $|\mathrm{GS}\rangle_\mathrm{AF}  =  \frac{1}{\sqrt{12}} \big( 2| \uparrow \downarrow \uparrow \downarrow\rangle + 2|\downarrow \uparrow \downarrow \uparrow\rangle - |\uparrow\uparrow \downarrow  \downarrow \rangle - |\uparrow \downarrow \downarrow \uparrow  \rangle  - |\downarrow \downarrow \uparrow \uparrow \rangle -|\downarrow \uparrow \uparrow \downarrow \rangle \big)$, ensuring lower energy $\langle \mathrm{GS}| \hat{H}_{\textrm{AF}} | \mathrm{GS} \rangle = 2J$ than the energy in the unentangeled  N\'{e}el state, $\langle \uparrow\downarrow \uparrow\downarrow | \hat{H}_{\textrm{AF}} | \uparrow\downarrow \uparrow\downarrow \rangle=J$. While the expectation values of localized spins in the  N\'{e}el state are $\langle \uparrow\downarrow \uparrow\downarrow | \hat{\mathbf{S}}_i| \uparrow\downarrow \uparrow\downarrow \rangle= \pm S\hbar$, in the true GS the same expectation values are  $_\mathrm{AF}\langle \mathrm{GS}  | \hat{\mathbf{S}}_i| \mathrm{GS} \rangle_\mathrm{AF} \equiv 0$ (akin to quantum spin liquids~\cite{Broholm2020}) for which no classical dynamics can be formulated.

In this study, we choose small ferromagnetic metal (FM) or antiferromagnetic metal (AFM) chains of $N=4$ spins $S=1$ or $S=5/2$. They interact via the NN Heisenberg exchange $J$ and are localized on lattice sites between which conduction electrons can hop while their movable spins interact with localized spins via $sd$ exchange coupling of strength $J_\mathrm{sd}$, as illustrated in Fig.~\ref{fig:fig1}. These chains encapsulate essential features of FM or AFM elements within spintronic devices, while their relatively small Hilbert space of dimension $4^N (2S+1)^N$ makes possible {\em numerically exact} calculations~\cite{Bajpai2021} of their quantum time-evolution.  

\begin{figure}[t]
	\centering
	\includegraphics[width=\linewidth]{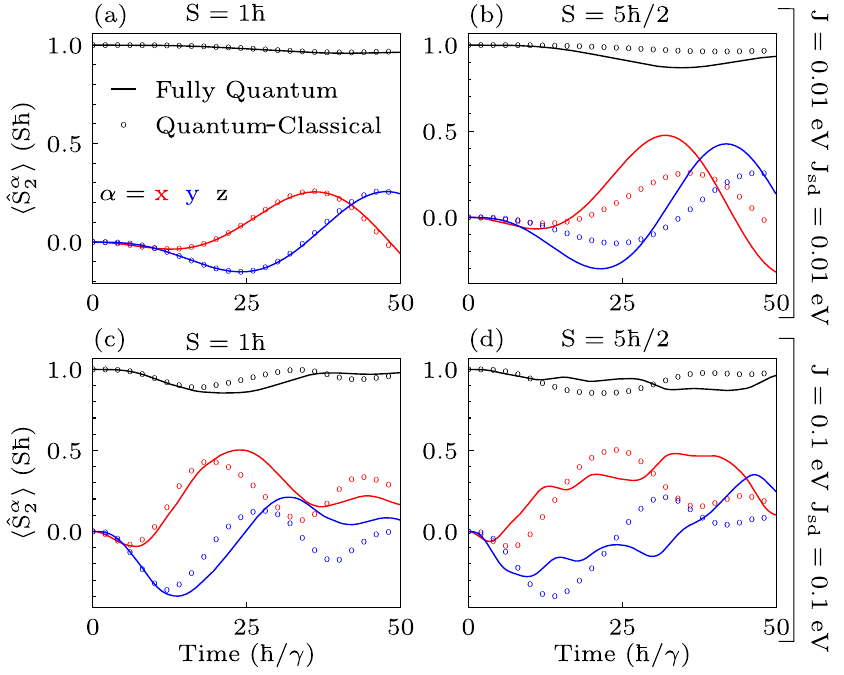}
	\caption{Time evolution of localized spin at site $i=2$ within FM chain of $N=4$ sites [Fig.~\ref{fig:fig1}(a)] from fully quantum (solid lines) vs. quantum-classical (open circles) calculations. Two different localized spins ($S=1$ and $S=5/2$) are used in fully quantum calculations, while quantum-classical calculations treat localized spins as unit vectors of fixed length $|\mathbf{S}_i|=1$. We also use two different sets of parameters [shown on the right for panels (a),(b) vs. panels (c),(d)] for Heisenberg and $sd$ exchange interaction between localized spins and electron spin and localized spins, respectively.}
	\label{fig:fig2}
\end{figure}

\begin{figure}[t]
	\centering
	\includegraphics[width=\linewidth]{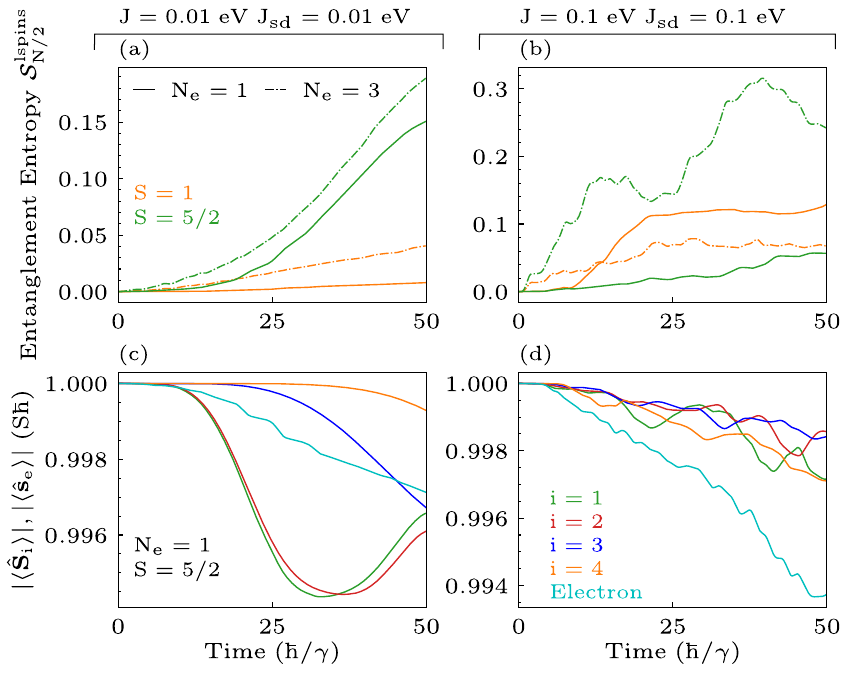}
	\caption{(a) and (b) Time evolution of the von Neumann entanglement entropy [Eq.~\eqref{eq:entropy}] for the subsystem of FM chain of $N=4$ sites  [Fig.~\ref{fig:fig1}(a)], composed of half~\cite{Petrovic2021a,Bardarson2012} of all localized spins (that is at sites $i=3,4$), which is extracted from fully quantum time evolution of the whole system of electrons and localized spins studied in Fig.~\ref{fig:fig2}. Cases with $N_e = 1$ (solid lines) or $N_e = 3$ (dash-dot lines) conduction electrons filling the chain are considered. Panels (c) and (d) show time evolution of the magnitude of the expectation values of localized spins and electron spin, in the case $S=5/2$ and $N_e = 1$, whose decay below one, such as $|\langle \hat{\mathbf{S}}_i \rangle(t)/S \hbar < 1$ and $|\langle \hat{\mathbf{s}}_e \rangle(t)2/\hbar < 1$, signifies growth of entanglement~\cite{Petrovic2021a,Bardarson2012} of such minimal subsystem to the rest of the total system.}
	\label{fig:fig3}
\end{figure}

\begin{figure*}
	\centering
	\includegraphics[width=\linewidth]{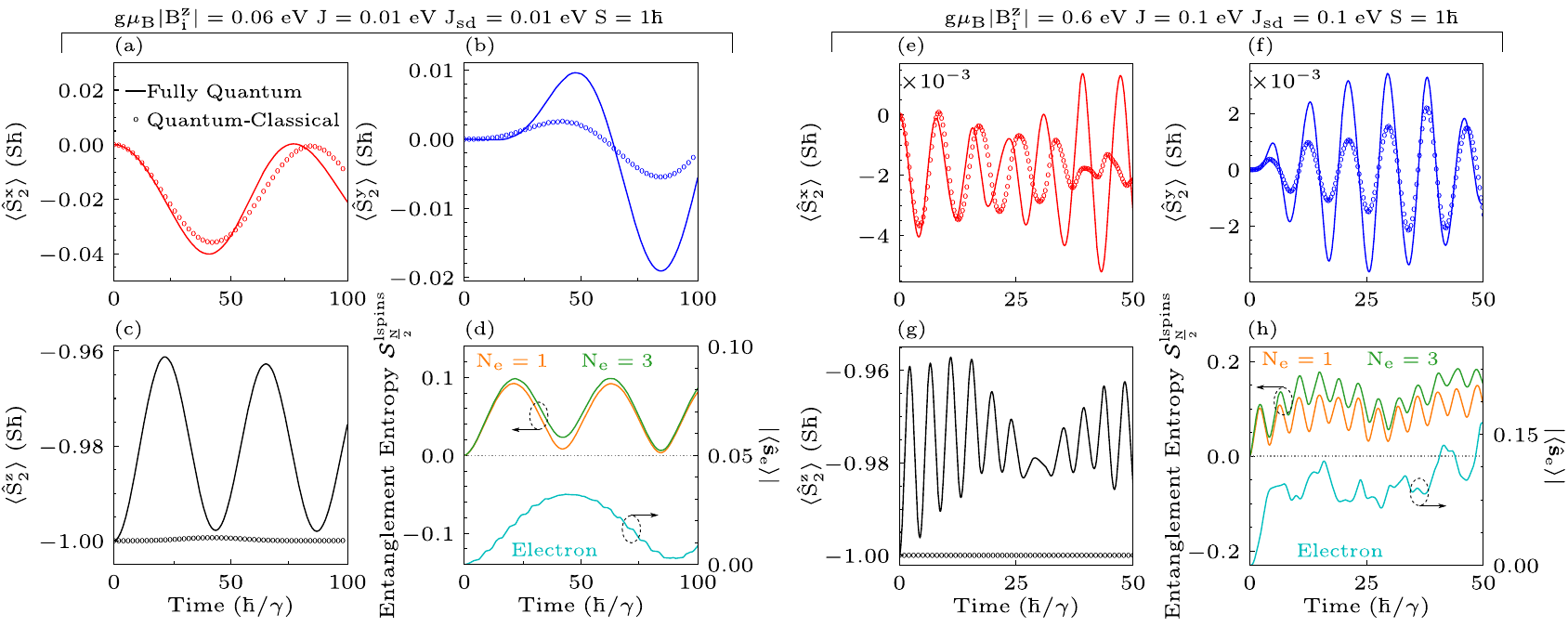}
	\caption{(a)--(d) Time evolution of localized spin at site \mbox{$i=2$}  within AFM chain of $N=4$ sites [Fig.~\ref{fig:fig1}(b)] from fully quantum (solid lines) vs. quantum-classical (open circles) calculations. The quantum localized spin is $S=1$ in fully quantum calculations, while quantum-classical calculations  treat localized spins as unit vectors of fixed length $|\mathbf{S}_i|=1$. Panel (d) shows the entanglement entropy for half ($i=3,4$) of all localized spins in the case with $N_e=1$ or $N_e=3$ electrons, as well as the magnitude of the expectation value of electron spin in the case $N_e=1$. Panels (e)--(h) are counterparts of panels (a)--(d) for larger values of exchange couplings $J_{sd}$ and $J$, as well as larger corresponding applied staggered magnetic field $B_i^z$ to ensure unentangled N\'{e}eel state as the ground state at $t \le 0$.}
	\label{fig:fig4}
\end{figure*}

The paper is organized as follows. Section~\ref{sec:methods} introduces model Hamiltonians, rigorous notation and useful concepts. Our principal results in Figs.~\ref{fig:fig2}--\ref{fig:fig4} compare quantum-classical trajectories $\mathbf{S}_i(t)$ to fully quantum trajectories $\langle \hat{\mathbf{S}}_i\rangle(t)$ in Sec.~\ref{sec:fm} for FM  [Fig.~\ref{fig:fig2}] and in Sec.~\ref{sec:afm}  for AFM [Fig.~\ref{fig:fig4}] cases while directly connecting the onset of deviation between them to dynamical buildup of entanglement [Figs.~\ref{fig:fig3} and \ref{fig:fig4}(d),(h)]. The impact of finite size and/or finite temperature on fully quantum trajectories is analyzed [Fig.~\ref{fig:fig5}--\ref{fig:fig7}] in Sec.~\ref{sec:temperature}. We conclude in Sec.~\ref{sec:conclusions}.

\section{Models and methods}\label{sec:methods}

The FM and AFM setups in Fig.~\ref{fig:fig1} are described by quantum many-body Hamiltonian of localized spins and conduction electrons
\begin{align}\label{eq:hamiltonian}
   \hat{H} &= \hat{H}_\mathrm{lspins} +  \hat{H}_e + \hat{H}_{e\mathrm{-lspins}}(t > 0) + \hat{H}_\mathrm{\bf B},
\end{align}
which acts in $\mathcal{H}_\mathrm{total} = \mathcal{F}_e \otimes \mathcal{H}_1\otimes\cdots\otimes\mathcal{H}_N$ as the tensor product of Fock space $\mathcal{F}_e$ for electrons and Hilbert spaces $\mathcal{H}_i$ for localized spins. The interaction between localized spins is captured by the quantum Heisenberg Hamiltonian 
\begin{equation}\label{eq:heisenberg}
\hat{H}_\mathrm{lspins} = - J\sum_{\langle ij \rangle}\hat{\bf S}_i\cdot\hat{\bf S}_j, 
\end{equation}
where $\langle ij \rangle$ indicates interaction between the NN sites.  Here the spin operator  
\begin{equation}\label{eq:spinop}
\hat{S}_i^\alpha = \underbrace{\mathbb{1}\otimes\mathbb{1}\cdots\mathbb{1}}_\text{$i-1$} \otimes \hat{S}^\alpha \underbrace{\otimes \mathbb{1}\otimes\cdots\otimes\mathbb{1}}_\text{$N-i$ times},
\end{equation}
at site $i$  is a matrix of size $(2S+1)^N$ for spin $S$,  where $\mathbb{1}$ is the unit matrix of size $(2S+1)$. The tight-binding Hamiltonian
\begin{equation}
\hat{H}_e = -\gamma \sum_{\langle ij\rangle,\sigma} \hat{c}^\dagger_{i\sigma} \hat{c}_{j\sigma},
\end{equation}
describes conduction electrons, where \mbox{$\gamma = 1$ eV} (setting the unit of energy) is the NN hopping between orbitals centered on each site $i$. Here $\hat{c}^\dagger_{i\sigma}$ ($\hat{c}_{i\sigma}$) creates (annihilates) an electron with spin $\sigma=\uparrow,\downarrow$ on site $i$, so that each site can host one of the four possible electronic states obeying the Pauli principle---empty $|0\rangle$, spin-up $\hat{c}^\dagger_{i\uparrow}|0\rangle$, spin-down $\hat{c}^\dagger_{i\downarrow}|0\rangle$, and doubly occupied  $\hat{c}^\dagger_{i\uparrow} \hat{c}^\dagger_{i\downarrow}|0\rangle$. The interaction between conduction electron spin and localized spins is taken into account by $sd$ Hamiltonian (in spintronics terminology~\cite{Ralph2008}; in other fields this term, together with first two terms in Eq.~\eqref{eq:hamiltonian}, is termed Kondo-Heisenberg model~\cite{Tsvelik2017})
\begin{equation}\label{eq:sd}
\hat{H}_{e\mathrm{-lspins}} = -J_\mathrm{sd}(t > 0) \sum_{i=1}^N \hat{\bf s}_i \cdot\hat{\bf S}_i,
\end{equation}
where \mbox{$\hat{s}^{\alpha}_{i}=\sum_{\sigma,\sigma'=\{\uparrow,\downarrow\}} \hat{c}^\dagger_{i\sigma} \hat{\sigma}^\alpha_{\sigma\sigma'} \hat{c}_{i\sigma'}$} is the local (per-site) spin density operators; $\hat{\sigma}^\alpha$ is the Pauli matrix; and \mbox{$\hat{s}^\alpha_e=\sum_{i=1}^N \hat{s}^\alpha_{i}$} is the operator of the total electron spin along the $\alpha$-axis ($\alpha \in \{x,y,z\}$). 

The last (Zeeman) term in Eq.~\eqref{eq:hamiltonian} is due to externally applied magnetic fields
\begin{equation}\label{eq:hb}
\hat{H}_\mathrm{\bf B} = -g\mu_B \sum_{i=1}^N B_i^z \hat{S}_i^z - g\mu_B B_1^x(t > 0) \hat{S}_1^x. 
\end{equation}
In the FM case  [Fig.~\ref{fig:fig1}(a)],  \mbox{$g \mu_B B_i^z = 0.1$ eV} ($\mu_B$ is the Bohr magneton) acts as anisotropy to select~\cite{Wieser2015} the positive $z$-axis as the easy axis, which is chosen to avoid usual $D_z(\hat{S}_i^z)^2$ breaking the correspondence between fully quantum and LL description from the outset~\cite{Wieser2011,Gauyacq2014} (see Sec.~\ref{sec:intro}). Additional \mbox{$g \mu_B B_1^x(t > 0) = 0.1$ eV} is applied at first site $i=1$ to initiate~\cite{Wieser2015} nonequilibrium dynamics for $t > 0$.  In the AFM case  [Fig.~\ref{fig:fig1}(b)], we use staggered field $B_i^z=-B_{i+1}^z$ of magnitude \mbox{$g \mu_B |B_i^z| = 0.06$ eV $> |J|$} or \mbox{$g \mu_B |B_i^z| = 0.6$ eV $> |J|$} for $|J|=0.01$ eV or $|J|=0.1$ eV, respectively. Such staggered field  enforces~\cite{Stoudenmire2012,Mitrofanov2021} the N\'{e}el state as the GS with zero entanglement at times $t \le 0$. Addition external magnetic field \mbox{$g \mu_B B_1^x(t > 0)=0.01$ eV} or \mbox{$g \mu_B B_1^x(t > 0)=0.1$ eV} for $|J|=0.01$ eV or $|J|=0.1$ eV, respectively,  is applied at the first site $i=1$ to initiates nonequilibrium dynamics of localized spins within the AFM setup [Fig.~\ref{fig:fig1}(b)]. 

In fully quantum calculations, the wavefunction of an isolated quantum many-body system electrons+localized-spins is evolved within $\mathcal{H}_\mathrm{total}$ using
\begin{equation}\label{eq:evolve}
|\Psi(t+\delta t)\rangle =  e^{-i\hat{H}\delta t/\hbar} |\Psi(t)\rangle,
\end{equation} 
where we employ the Crank-Nicolson algorithm~\cite{Wells2019} with time step \mbox{$\delta t = 0.1$ $\hbar/\gamma$}. The initial state 
\begin{equation}\label{eq:psi0}
|\Psi(t=0) \rangle \equiv  |\chi(t=0)\rangle_e \otimes |\Sigma(t=0) \rangle_\mathrm{lspins},
\end{equation}
is unentangled (or separable) GS, whose factor $|\Sigma \rangle_\mathrm{lspins}$ for localized spins is also unentangled---\mbox{$|\Sigma(t=0) \rangle_\mathrm{lspins}=|\uparrow \uparrow \ldots \uparrow \uparrow \rangle$} for FM or \mbox{$|\Sigma(t=0) \rangle^{\rm N\acute{e}el}_\mathrm{lspins}=|\uparrow \downarrow \ldots \uparrow \downarrow \rangle$}---as a prerequisite~\cite{Wieser2015} for comparing quantum vs. classical dynamics of localized spins. The unentangled GS is ensured by switching on $\hat{H}_{e\mathrm{-lspins}}(t > 0)$ in Eq.~\eqref{eq:hamiltonian} for times $t > 0$, otherwise GS would be entangled~\cite{Bajpai2021} for $t \le 0$ with entanglement entropy $\mathcal{S}^\mathrm{lspins}_N=\mathcal{S}_e \simeq 10^{-3}$. 

The von Neumann entanglement entropy~\cite{Chiara2018,Bardarson2012,Petrovic2021a} of a chosen subsystem  is computed as 
\begin{equation}\label{eq:entropy}
\mathcal{S}_\mathrm{sub}(t) = -\text{Tr}\big[ \hat{\rho}(t)_\text{sub} \ln \hat{\rho}(t)_\text{sub} \big], 
\end{equation}
using the reduced density matrix 
\begin{equation}\label{eq:reduceddm}
\hat{\rho}_\text{sub}(t) = \text{Tr}_\text{other} |\Psi(t) \rangle \langle \Psi(t)|, 
\end{equation}
obtained by partial trace of pure state density matrix over all other degrees of freedom. Note  that when total system is split into two parts, they always have identical entropies, such as $\mathcal{S}^\mathrm{lspins}_N=\mathcal{S}_e$.

In quantum-classical calculations, we switch to classical Hamiltonian
\begin{equation}\label{eq:hclassical}
H_\mathrm{lspins} = - J \sum_{\langle ij\rangle} {\bf S}_i\cdot{\bf S}_j - J_\text{sd} \sum_{i = 1}^N \langle \hat{\bf s}_i \rangle(t) \cdot{\bf S}_i + {H}_{\bf B}, 
\end{equation}
for localized spins $\mathbf{S}_i$ where 
\begin{equation}\label{eq:hbclassical}
H_\mathrm{\bf B} = -g\mu_B \sum_{i=1}^N B_i^z S_i^z - g\mu_B B_1^x(t > 0) S_1^x,
\end{equation}
is the classical version of $\hat{H}_{\bf B}$ in Eq.~\eqref{eq:hamiltonian}. The quantum Hamiltonian for electrons then   retains only two terms from $\hat{H}$ in Eq.~\eqref{eq:hamiltonian} 
\begin{equation}\label{eq:hquantum}
\hat{H}(t) =  \hat{H}_e -J_\mathrm{sd} \sum_{i= 1}^N \hat{\bf s}_i \cdot {\bf S}_i (t), 
\end{equation}
with its time-dependence being generated by self-consistent coupling~\cite{Petrovic2018,Sayad2015,Elbracht2020} to the LL equation. In order 
to compare directly expectation values of the quantum mechanical localized spins $\langle \hat{S}_i \rangle (t)/S \hbar$ and classical localized spins, ${\bf S}_i(t)$ (both types of vectors have unit modulus at $t=0$),  we set~\cite{Wieser2011} $\hbar = g = 1$.  

In addition, we set the Gilbert damping $\lambda=0$ in Eq.~\eqref{eq:llg}, as also used in previous comparisons~\cite{Wieser2011,Wieser2015,Wieser2013,Gauyacq2014} of quantum vs. classical dynamics for localized spins in the absence of electrons. This choice is due to their quantum calculations handling F insulator, or ours handling F or AF metal, which are considered as isolated quantum systems that are not coupled to natural intrinsic bosonic baths (such as phonons of the crystal lattice or photons of the ambient electromagnetic environment), external baths~\cite{Elbracht2020} or external fermionic reservoirs~\cite{Petrovic2018,Bajpai2019a,Bajpai2020,Petrovic2021b,Bostrom2019}. All such environments would contain an infinite number of degrees of freedom, therefore supplying continuous energy spectrum that ensures dissipation in problems of quantum statistical mechanics. 

Nevertheless, our calculations are still qualitatively different from those of Refs.~\cite{Wieser2011,Wieser2015,Wieser2013,Gauyacq2014} because excitation energy and excess spin can be transported away~\cite{Bajpai2019a,Sayad2015} from localized spins by {\em nonequilibrium} conduction electrons driven out of equilibrium by time dependence of $\mathbf{S}_i(t)$. While this effect is often modeled by adding spin-transfer torque (STT), $\frac{g}{\mu_M}\mathbf{T}_i\left[\mathbf{S}_i(t)\right]$, into the LL equation as electronic back-action~\cite{Bajpai2020,Petrovic2021b,Zhang2004,Zhang2009,Tatara2019}, in our quantum-classical calculations it is automatically included by self-consistent loop where such term can be nonlocal in space~\cite{Petrovic2021b,Bajpai2020} and time~\cite{Bajpai2019a,Sayad2015}. Besides conventional  Slonczewski-Berger~\cite{Berkov2008,Ralph2008} STT \mbox{$\mathbf{T}_i\left[\mathbf{S}_i(t)\right] \propto \langle \hat{\bf s}_i \rangle(t) \times \mathbf{S}_i(t)$}, our fully quantum calculation also include {\em quantum STT}~\cite{Petrovic2021,Petrovic2021a,Mitrofanov2021,Zholud2017} where transfer of spin angular momentum between electrons and localized spins is possible even when $\langle \hat{\bf s}_i \rangle$ is {\em collinear but antiparallel}~\cite{Petrovic2021a} to the direction of $\langle \hat{\mathbf{S}}_i \rangle$.


\section{Results and discussion}\label{sec:results}

\begin{figure*}
	\centering
	\includegraphics[width=\linewidth]{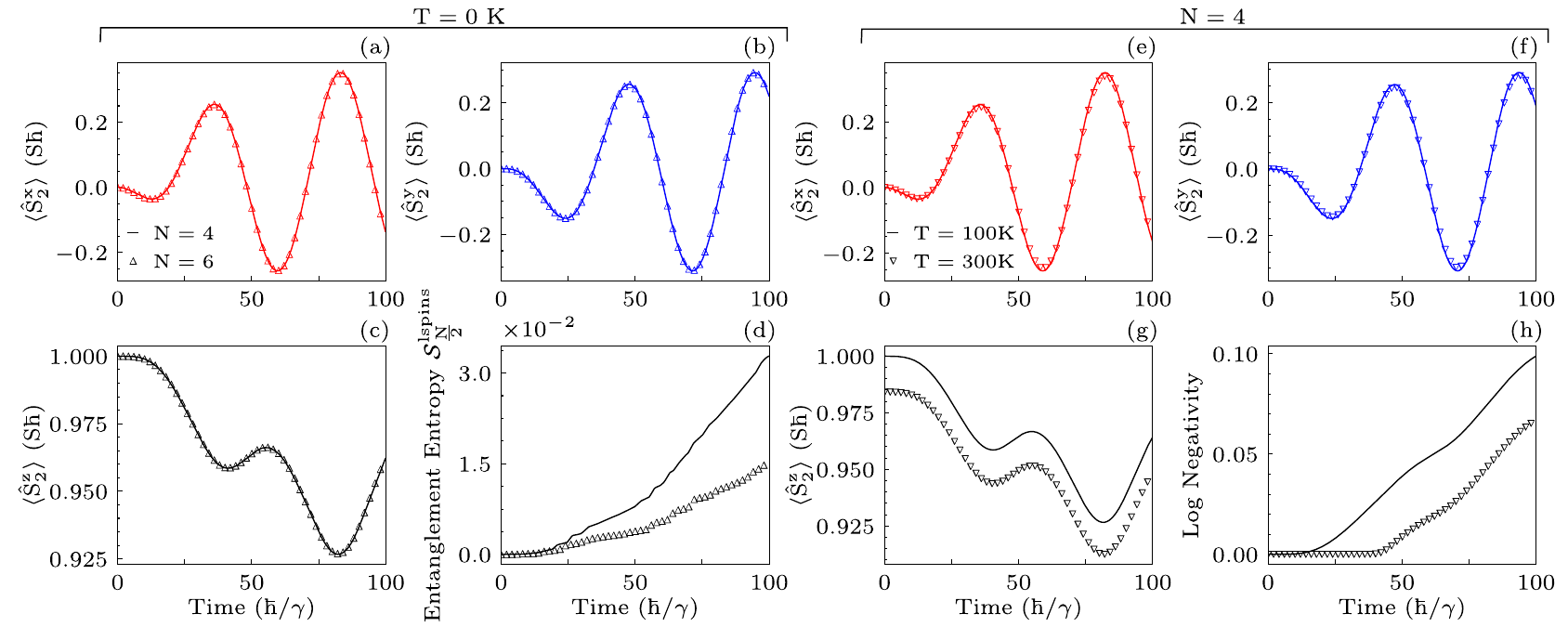}
	\caption{(a)--(c) Time evolution of localized spin at site \mbox{$i=2$} within FM chain of $N=4$ sites (solid lines) or $N=6$ (up triangles) from fully quantum calculations at zero temperature \mbox{$T=0$ K}. Panel (d) shows time evolution of the von Neumann entanglement entropy [Eq.~\eqref{eq:entropy}] for half ($i=3,4$) of all localized spins. (e)--(g) Time evolution of localized spin at site \mbox{$i=2$} within FM chain of $N=4$ sites at \mbox{$T=100$ K} (solid lines) or \mbox{$T=300$ K} (down triangles) from fully quantum calculations. Panel (d) shows time evolution of mutual logarithmic negativity $E_\mathrm{Neg}(1,2|3,4)$ [Eq.~\eqref{eq:negativity}]  between localized spins at sites $i=1,2$ and those at sites $i=3,4$. In all panels we use $S=1$ localized spins; $J=0.01$ eV; $J_\mathrm{sd}=0.01$ eV; and $N_e=1$ conduction electron.}
	\label{fig:fig5}
\end{figure*}

\subsection{Fully quantum vs. quantum-classical dynamics in ferromagnetic metal}\label{sec:fm}
Let us first recall that electronic hopping $\gamma$ and $sd$ exchange $J_\mathrm{sd}$ control the speed of electron translational and spin dynamics, respectively. For example, large ratio $J_\mathrm{sd}/\gamma$ is required to bring the system of conduction electrons and localized spins into the ``adiabatic limit''~\cite{Petrovic2021b,Tatara2019} where electron spin remains in the lowest energy state at each time while locking to the direction of localized spins.  Figure~\ref{fig:fig2}(a),(b) shows that small $J$ and $J_\mathrm{sd}$ make possible for quantum-classical trajectories $\mathbf{S}_i(t)$ to re-trace fully quantum trajectories $\langle \hat{\mathbf{S}}_i\rangle(t)$ in the FM case [Fig.~\ref{fig:fig1}(a)] over long time intervals, but this correspondence is lost by increasing $J$ and $J_\mathrm{sd}$ [Fig.~\ref{fig:fig2}(c),(d)]. Surprisingly, in contrast to the na\"{i}ve~\cite{Goldberg2020} ``rule of thumb''~\cite{Berkov2008,Kim2010,Evans2014,Parkinson1985},  as well as  previous rigorous comparisons~\cite{Wieser2011,Wieser2015,Wieser2013,Gauyacq2014} of quantum vs. classical trajectories of localized spins in the {\em absence} of conduction 
electrons---where  $S \rightarrow \infty$ justifies applicability of the LL equation---increasing localized spins from $S=1$ [Fig.~\ref{fig:fig2}(a),(c)] to $S=5/2$ [Fig.~\ref{fig:fig2}(b),(d)] leads to {\em larger deviations} between $\mathbf{S}_i(t)$ and  $\langle \hat{\mathbf{S}}_i\rangle(t)$. 

We explain the emergence of such large deviations in the course of time evolution by dynamical buildup of entanglement entropy of 
localized spins subsystem [Fig.~\ref{fig:fig3}]. The buildup is facilitated by increasing $J_\mathrm{sd}$ that fosters entangling localized spins to conduction electrons; or increasing $S$, which enlarges the environment to which individual [Fig.~\ref{fig:fig3}(c),(d)] localized spins  or electronic spin can entangle. As soon as entanglement becomes nonzero at some onset time $t>0$, the magnitude $|\langle \hat{\mathbf{S}}_i\rangle(t) | < S\hbar$ starts to shrink  [Fig.~\ref{fig:fig3}(c),(d)] which makes the LL equation {\em inapplicable}~\cite{Wieser2015} because such effect is forbidden in any classical dynamics.

\subsection{Fully quantum vs. quantum-classical dynamics in antiferromagnetic metal}\label{sec:afm}
In the AFM case [Fig.~\ref{fig:fig1}(b)], large deviations in Fig.~\ref{fig:fig4}(a)--(c) between $\mathbf{S}_i(t)$ and  $\langle \hat{\mathbf{S}}_i\rangle(t)$ are present from the outset $t \rightarrow 0+$, despite using: ({\em i}) unentangled N\'{e}el state $|\Sigma \rangle^{\rm N\acute{e}el}_\mathrm{lspins}$ as the initial condition at $t=0$; and ({\em ii}) small values $S=1$, $J$ and $J_\mathrm{sd}$ [Fig.~\ref{fig:fig4}(a)--(d)] 
which in the case of FM chain {\em ensure} [Fig.~\ref{fig:fig2}(a)] good match between quantum-classical and fully quantum  trajectories. This is explained by much more rapid increase of entanglement entropy of localized spins at $t \rightarrow 0+$ in the AFM case [Fig.~\ref{fig:fig4}(d) and ~\ref{fig:fig4}(h)] than in the FM case [Figs.~\ref{fig:fig3}(a),(b)].

\begin{figure*}
	\centering
	\includegraphics[width=\linewidth]{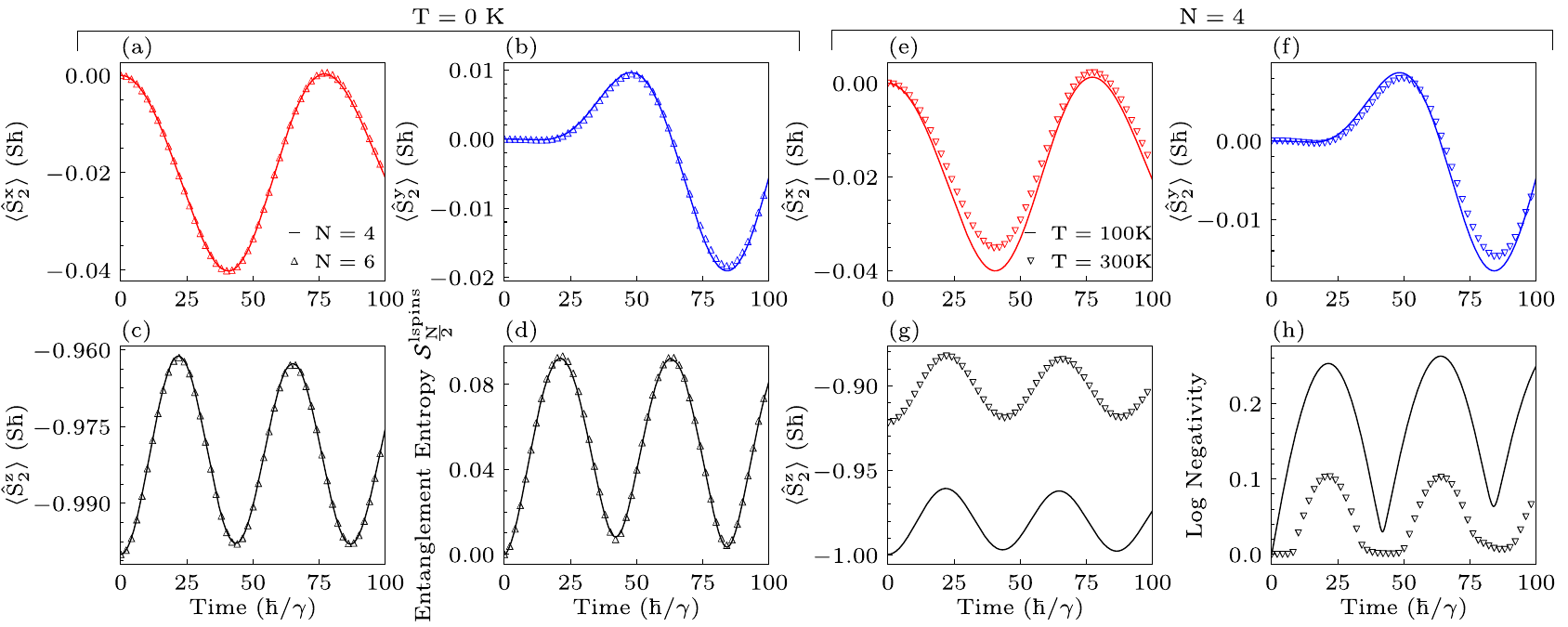}
	\caption{(a)--(c) Time evolution of localized spin at site \mbox{$i=2$} within AFM chain of $N=4$ sites (solid lines) or $N=6$ (up triangles) from fully quantum calculations at zero temperature \mbox{$T=0$ K}. Panel (d) shows time evolution of the von Neumann entanglement entropy [Eq.~\eqref{eq:entropy}] for half ($i=3,4$) of all localized spins in the case. (e)--(g) Time evolution of localized spin at site \mbox{$i=2$} within FM chain of $N=4$ sites at \mbox{$T=100$ K} (solid lines) or \mbox{$T=300$ K} (down triangles) from fully quantum calculations. Panel (d) shows time evolution of mutual logarithmic negativity $E_\mathrm{Neg}(1,2|3,4)$ [Eq.~\eqref{eq:negativity}] between localized spins at sites $i=1,2$ and those at sites $i=3,4$. In all panels we use $S=1$ localized spins; $J=0.01$ eV; $J_\mathrm{sd}=0.01$ eV; staggered magnetic field $g\mu_B|B_i^z|=0.06$ eV; and $N_e=1$ conduction electron.}
	\label{fig:fig6}
\end{figure*}

Note that in both FM and AFM cases, increasing the number of conduction electrons from $N_e=1$, where a single spin-up electron is placed on site $i=1$
\begin{equation}\label{eq:einitial1}
|\chi(t=0) \rangle_e = \hat{c}^\dagger_{1\uparrow}|0\rangle,
\end{equation}
to $N_e=3$, where spin-up and spin-down electrons are placed on site $i=1$ and one spin-up electron is placed on site 2 
\begin{equation}\label{eq:einitial3}
|\chi(t=0) \rangle_e = \hat{c}^\dagger_{1\uparrow} \hat{c}^\dagger_{1\downarrow} \hat{c}^\dagger_{2\uparrow}|0\rangle,
\end{equation} 
leads to similar entanglement entropies [Figs.~\ref{fig:fig3}(a),(b) and ~\ref{fig:fig4}(d),(h)] and trajectories (not shown explicitly) of $\langle \hat{\mathbf{S}}_i\rangle(t)$ because $|\langle \hat{\mathbf{s}}_e \rangle(t=0)| = \hbar/2$ in both $N_e=1$ and $N_e=3$ cases. 

\subsection{Finite size and finite temperature effects in fully quantum dynamics}\label{sec:temperature}

\begin{figure}
	\centering
	\includegraphics[width=\linewidth]{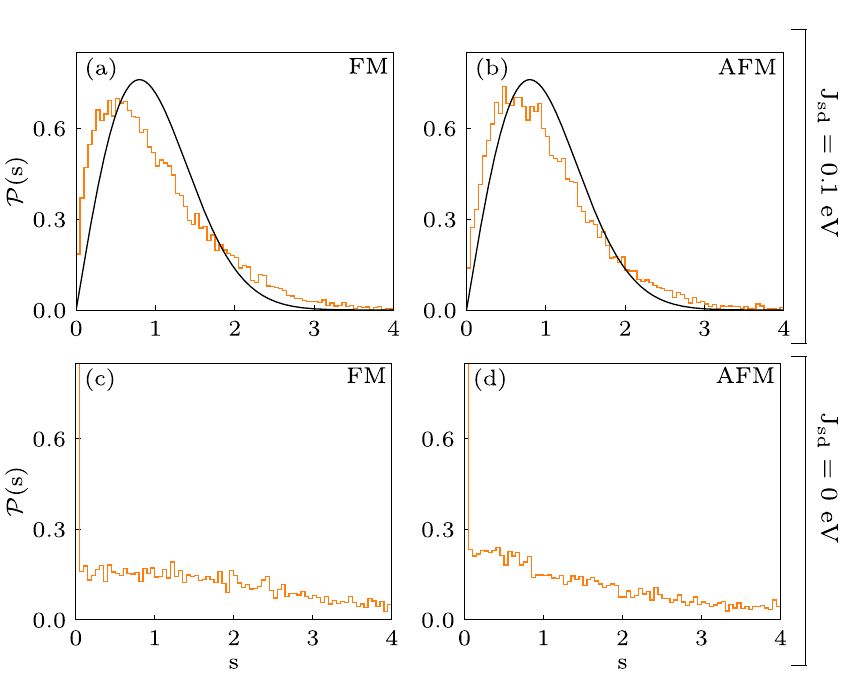}
	\caption{Energy level-spacing statistics $\mathcal{P}(s)$ (orange curve histogram), after the unfolding procedure (Sec.~\ref{sec:temperature}), for: (a),(c) FM or (b),(d) AFM setups in Fig.~\ref{fig:fig1} composed of $N=4$ sites hosting localized spins $S=5/2$. Smooth black lines in panels (a) and (b) are the WD distribution $\mathcal{P}_\mathrm{WD}^\mathrm{GOE}$ [Eq.~\eqref{eq:wd}]. In panels (a),(b) we use $J_\mathrm{sd}=0.1$ eV and $N_e=1$ conduction electron, while in panels (c),(d) the interaction between electron spin and localized spin is {\em switched off}, $J_\mathrm{sd}=0$. The applied homogeneous magnetic field is \mbox{$g\mu_BB_i^z=0.1$ eV} in (a),(c), and the applied staggered magnetic field in (b),(d) is \mbox{$g\mu_B|B_i^z|=0.6$ eV}.}
	\label{fig:fig7}
\end{figure}

To examine how sensitive our results are on the system size, Figs.~\ref{fig:fig5}(a)--(d) for FM and ~\ref{fig:fig6}(a)--(d) for AFM repeat fully quantum  calculations from Figs.~\ref{fig:fig2} and ~\ref{fig:fig4}, respectively, using slightly larger chains composed of $N=6$ sites hosting $S=1$ localized spins. No significant changes in the dynamics of localized spins are observed when increasing the chain length from $N=4$ to $N=6$ for the time interval considered. This suggests that results for a system size of just $N=4$ sites could already provide a good approximation to the thermodynamic limit, at least in this particular problem. Similar observation has been made in field-theoretic studies~\cite{Schuckert2018} of specific problems in the dynamics of quantum spin chains in the absence of electrons (see, e.g., Fig.~\ref{fig:fig7} in Ref.~\cite{Schuckert2018} comparing localized spin trajectories for chains of length $N=4,6,10$).

We also demonstrate that even small FM [Fig.~\ref{fig:fig7}(a)] or AFM  [Fig.~\ref{fig:fig7}(b)] chains employed in our study, with $N=4$ sites whose localized spin $S=5/2$ interact with a single $N_e=1$ conduction electron, are sufficient to exhibit energy level repulsion leading to quantum-chaotic energy-level-statistics~\cite{Gubin2012,Allesio2016} where the level spacing $\mathcal{P}(s)$ follows the Wigner-Dyson (WD) distribution [smooth black lines in Fig.~\ref{fig:fig7}(a),(b)]
\begin{equation}\label{eq:wd}
	\mathcal{P}_\mathrm{WD}^\mathrm{GOE}(s) = \frac{\pi s}{2} e^{-\pi s^2/4}.
\end{equation}
Here the WD distribution~\cite{Gubin2012,Allesio2016} is written for Gaussian orthogonal ensemble (GOE) in random matrix theory because Hamiltonian in Eq.~\eqref{eq:hamiltonian} has real symmetric matrix representation. The distribution $\mathcal{P}(s)$ is obtained for unfolded (i.e., mean level spacing is one~\cite{Gubin2012,Allesio2016})  neighboring energy-level spacing $s_n$. Following Ref.~\cite{Jansen2019}, we start from $\delta_n = E_{n+1}-E_n$; introduce a cumulative spectral function $G(E) = \sum_n \Theta(E-E_n)$, where $\Theta$ is the unit step function; and we additionally smoothen $G(E)$ by fitting it with a polynomial of degree six $g_6(E)$. The spectral analysis is then performed on the unfolded neighboring level spacings $s_n = g_6(E_{n+1}) - g_6(E_n)$. For this unfolding procedure we use 66\% of energy eigenvalues in the middle of the spectrum. 

Let us recall that isolated quantum many-body systems, with quantum-chaotic statistics of energy levels~\cite{Gubin2012,Allesio2016}, can exhibit thermalization without invoking any external bath. That is, despite unitary time evolution of isolated quantum systems from an initial nonequilibrium state, as is the case of evolution presented in Figs.~\ref{fig:fig2}--\ref{fig:fig4}, they can thermalize (that is, local observables relax towards
a Gibbs ensemble with an effective temperature) at the level of individual eigenstates if the stringent conditions of the so-called eigenstate thermalization hypothesis (ETH)~\cite{Allesio2016} are satisfied. Note that quantum-chaotic level-spacing statistics is prerequisite for ETH~\cite{Allesio2016}, but it cannot distinguish between completely ergodic system and a system with a nonzero number of nonergodic eigenstates whose fraction vanishes in the thermodynamic limit. The interaction between conduction electron spin and localized spins plays a {\em crucial role} to generate quantum-chaotic level-spacing statistics in Fig.~\ref{fig:fig7}(a),(b) which is absent when we set $J_\mathrm{sd}=0$ in  Fig.~\ref{fig:fig7}(c),(d). The role of interaction with even a single electron in  generating quantum-chaotic level-statistics has been noted in other types of (small) quantum many-body systems, such as those hosting phonons~\cite{Jansen2019} instead of localized spins in the focus of our study. 

Finally, we couple FM or AFM systems in Fig.~\ref{fig:fig1} to a real macroscopic external bath at temperature $T$ [Figs.~\ref{fig:fig5}(e)--(h) and Fig.~\ref{fig:fig6}(e)--(h)] in order to examine whether thermal fluctuations can suppress dynamical buildup of entanglement, taking into account that such suppression is the {\em key} prerequisite [Secs.~\ref{sec:fm} and ~\ref{sec:afm}] for possible transition of fully quantum into classical dynamics of localized spins. For this purpose, we replace pure initial state in Eq.~\eqref{eq:psi0} with the equilibrium density matrix $\hat{\rho}_\mathrm{eq}$ in the canonical ensemble 
\begin{equation}\label{eq:rhoeq}
\hat{\rho}(t=0) = \hat{\rho}_\mathrm{eq} = \frac{1}{Z}e^{-\hat{H}/k_BT} = \hat{\rho}_e \otimes \hat{\rho}_1 \otimes \hat{\rho}_2 \otimes \hat{\rho}_3 \otimes \hat{\rho}_4,
\end{equation} 
where $Z=\mathrm{Tr}\, e^{-\hat{H}/k_BT}$ is the partition function. This {\em mixed} quantum state of $N_e=1$ electron and $N=4$ localized spins is classified  as {\em separable or unentangled}~\cite{Chiara2018,Wu2020,Elben2020a,Sang2021}, which is ensured by using $J_\mathrm{sd}(t \le 0)=0$ and staggered magnetic field in the AFM case in Hamiltonian [Eq.~\eqref{eq:hamiltonian}]. With this initial condition, the unitary time evolution is generated via the von Neumann equation
\begin{equation}\label{eq:vn}
i\hbar \frac{\partial \hat{\rho}}{\partial t} = \left[\hat{H}, \hat{\rho} \right], 
\end{equation}
which is solved by using the Runge$-$Kutta 4th order method with time step \mbox{$\delta t = 0.1$ $\hbar/\gamma$}. 

Because $\hat{\rho}(t=0)=\hat{\rho}_\mathrm{eq}$ in Eq.~\eqref{eq:rhoeq} has nonzero von Neumann entropy, we switch to mutual {\em logarithmic negativity}~\cite{Chiara2018}
\begin{eqnarray}\label{eq:negativity}
	E_\mathrm{Neg}(1,2|3,4) & \equiv & E_\mathrm{Neg}(\hat{\rho}_{1,2 \cup 3,4}) = \ln || \hat{\rho}^{T_{1,2}}_{1,2 \cup 3,4} ||   \nonumber \\
	&=& \ln || \hat{\rho}^{T_{1,2}}_{1,2 \cup 3,4} || = \ln \sum_n |\lambda_n|,
\end{eqnarray}
as the measure~\cite{Wu2020,Elben2020a,Sang2021}  of entanglement in mixed quantum state $\hat{\rho}(t)$. This quantity is defined using subsystems composed of localized spins $1$ and $2$ or of localized spins $3$ and $4$. Here $||\hat{A}|| = \mathrm{Tr}|\hat{A}|=\mathrm{Tr} \sqrt{\hat{A}^\dagger \hat{A}}$ is the trace norm of operator $\hat{A}$, and $\lambda_n$ are the eigenvalues of $\hat{\rho}^{T_{1,2}}_{1,2 \cup 3,4}$ or $\hat{\rho}^{T_{3,4}}_{1,2 \cup 3,4}$. 
The matrix elements of the partial transpose with respect to, e.g., subsystem composed of localized spins $1$ and $2$ are given by~\cite{Chiara2018}
\begin{equation}\label{eq:partialtranspose}
	\big(\hat{\rho}^{T_{1,2}}_{1,2 \cup 3,4} \big)_{i\alpha;j\beta} = \big(\hat{\rho}_{1,2 \cup 3,4} \big)_{j\alpha;i\beta}
\end{equation}
using matrix elements of the reduce density matrix of all $N=4$ localized spins, $\hat{\rho}_{1,2 \cup 3,4}$,
\begin{equation}\label{eq:matrixelements}
	\big(\hat{\rho}_{1,2 \cup 3,4} \big)_{i\alpha; j\beta} = {}_{1,2} \langle i| \, {}_{3,4}\langle \alpha|\hat{\rho}_{1,2 \cup 3,4}| j \rangle_{1,2} |\beta \rangle_{3,4}.
\end{equation}
In contrast to nonzero von Neumann entropy of unentangled mixed quantum state in Eq.~\eqref{eq:rhoeq}, logarithmic negativity is {\em zero} for $\hat{\rho}(t=0)=\hat{\rho}_\mathrm{eq}$, but it increases at later times in Figs.~\ref{fig:fig5}(h) and ~\ref{fig:fig6}(h) which necessarily implies dynamical buildup of entanglement between localized spins. That is, their finite-temperature mixed quantum state out of equilibrium cannot be written as a product state, $\rho_{1,2 \cup 3,4}(t) \neq \hat{\rho}_1(t) \otimes \hat{\rho}_2(t) \otimes \hat{\rho}_3(t) \otimes \hat{\rho}_4(t)$, which makes the LL equation at finite temperature (i.e., Eq.~\eqref{eq:llg} with an additional noise term~\cite{Garanin2017}) inapplicable to localized spins. Most importantly, increasing temperature does delay the onset of dynamical buildup of nonzero entanglement [compare solid lines with down triangles in Figs.~\ref{fig:fig5}(h) and ~\ref{fig:fig6}(h)], but it does not completely suppresses it [down triangles in Figs.~\ref{fig:fig5}(h) and ~\ref{fig:fig6}(h)] even at room temperature \mbox{$T=300$ K}.

\section{Conclusions and outlook}\label{sec:conclusions}
The quantum-classical and fully quantum-many-body computed trajectories of localized spins in FM, where they interact with conduction electrons, can be matched in some parameter regimes and time intervals [Figs.~\ref{fig:fig2} and ~\ref{fig:fig3}], akin to previous  comparisons~\cite{Wieser2011,Wieser2015,Wieser2013,Gauyacq2014} focused on F insulators {\em without} conduction electrons. However, in sharp contrast to these  studies, two types of trajectories deviate more (rather than less as in Refs.~\cite{Wieser2011,Wieser2015,Wieser2013,Gauyacq2014}) from each other with increasing value of localized spins $S$ [Fig.~\ref{fig:fig2}(b),(d)]. This is explained by dynamical buildup of larger entanglement with increasing $S$ [Figs.~\ref{fig:fig3}(a),(b)], which then precludes transition of fully quantum into quantum-classical trajectories. 

Furthermore, no match whatsoever is found in the case of AFM, which casts a doubt on applicability of phenomenological theories~\cite{Hals2011} [based on the LL Eq.~\eqref{eq:llg}] widely used in antiferromagnetic spintronics~\cite{Baltz2018}. We note that very recent neutron scattering experiments~\cite{Mathew2020,Scheie2021} have succeeded to directly detect and, furthermore, quantify  entanglement of macroscopically large number of localized spins within quasi-1D quantum antiferromagnets at finite temperature in equilibrium. This was achieved by measuring quantum Fisher information, as a witness for genuinely multipartite entanglement~\cite{Scheie2021}, which was extracted from direct measurements of dynamic susceptibility~\cite{Hauke2016,Brukner2006}. The measured multipartite entanglement vanishes for \mbox{$T \gtrsim 50$ K}. On the other hand, in our calculations for AFM case [Figs.~\ref{fig:fig4} and ~\ref{fig:fig6}] we use initial state with zero entanglement in equilibrium and unravel {\em dynamically} generated entanglement out of equilibrium  where simply increasing  temperature can postpone the onset of its growth [Fig.~\ref{fig:fig6}(h)] but it cannot completely suppresses it. 

Thus, we conclude that thermal fluctuations alone are not sufficient to ensure transition from quantum many-body entangled to quantum-classical nonequilibrium states with the LL equation being applicable to the latter. Instead, one has to  identify (which we relegate for future studies) specific decoherence channels due to the environment external to both electronic and localized spins subsystems, which can disrupt superpositions of quantum many-body states and maintain entanglement entropy of localized spins [Figs.~\ref{fig:fig3}(a),(b); ~\ref{fig:fig4}(d),(h); \ref{fig:fig5}(h); and ~\ref{fig:fig6}(h)] {\em as close as possible to zero} for all times so that nonclassical effects, which the LL equation can never capture, are {\em suppressed}.

\begin{acknowledgments}
This work was supported by the U.S. National Science Foundation (NSF) Grant No. ECCS 1922689. 
\end{acknowledgments}



\end{document}